
\input harvmac.tex

\noblackbox

\def\boxit#1{\vbox{\hrule\hbox{

\vrule\kern3pt\vbox{\kern3pt#1\kern3pt}\kern3pt\vrule}\hrule}}




%

\Title{\vbox{\baselineskip12pt
\hbox{DAMTP/R 94-52}
\hbox{UCSBTH-94-37}
\hbox{gr-qc/9501014}}}
{\vbox{\centerline{The Gravitational Hamiltonian, Action,} \centerline {Entropy
 and Surface Terms}}}

\baselineskip=12pt

\centerline{S. W. Hawking}

\medskip

\centerline{ \it
Department of Applied Mathematics and Theoretical Physics}
\centerline{\it Silver St.,
Cambridge CB3 9EW}
\centerline{\it Internet: swh1@amtp.cam.ac.uk}

\bigskip

\centerline{Gary T. Horowitz}

\medskip

\centerline{\it Physics Department}

\centerline{\it University of California}

\centerline{\it Santa Barbara, CA. 93111}

\centerline{\it Internet: gary@cosmic.physics.ucsb.edu }

\bigskip

\centerline{\bf Abstract}

\medskip

We give a general derivation of the gravitational hamiltonian
starting from the Einstein-Hilbert action, keeping track of all
surface terms. The surface term that arises in the hamiltonian can be taken
as the definition of the
`total energy', even for spacetimes that are not asymptotically
flat. (In the asymptotically flat case, it agrees with the usual ADM energy.)
We also discuss the relation between the euclidean action and the hamiltonian
when there are horizons of infinite area
(e.g. acceleration
horizons) as well as the usual finite area black hole horizons.
Acceleration horizons seem to be more analogous to extreme than
nonextreme black holes, since we find evidence that their horizon area is
not related to the total entropy.


\Date{January 1995}

\def\b{\beta}

\def\p{\partial}
\def\r{\rho}
\def\d{\nabla}
\def\vp{\varphi}

\def\H{{\cal H}}
\def\I{\tilde I}

\def\R{{\cal R}}
\def\S{\Sigma}
\def\RN{Reissner-Nordstr\"om}
\def\vp{\varphi}
\def\({\left (}
\def\){\right )}
\def\[{\left [}
\def\]{\right ]}

\def\np {  Nucl. Phys. }
\def \pl { Phys. Lett. }

\def \prl { Phys. Rev. Lett. }
\def \pr  { Phys. Rev. }
\def \cqg { Class. Quantum Grav.}
\def \cmp {Commun. Math. Phys.}
\gdef \jnl#1, #2, #3, 1#4#5#6{ {\sl #1~}{\bf #2} (1#4#5#6) #3}

\lref\hr{ S.  W.  Hawking and S.  R.  Ross,
``Duality Between Electric and Magnetic Black Holes", paper
in preparation}
\lref\hawking{ S. W. Hawking, \jnl \cmp, 43, 199, 1975.}
\lref\hawk{S. W. Hawking, in {\it General Relativity, an Einstein Centenary
Survey}, eds. S. W. Hawking and W. Israel (Cambridge University Press) 1979.}
\lref\hhr{ S. W. Hawking, G. T. Horowitz, and S. R. Ross, ``Entropy, Area, and
Black Hole Pairs", DAMTP/R 94-26, UCSBTH-94-25, gr-qc/9409013.}
\lref\ggs{D. Garfinkle, S.B. Giddings and A. Strominger, \jnl \pr,
D49, 958, 1994.}
\lref\gh{G.W.  Gibbons and S.W.  Hawking, \jnl \pr, D15, 2752, 1977.}
\lref\dgkt{H.F. Dowker, J.P. Gauntlett, D.A. Kastor and J. Traschen,
\jnl \pr,  D49, 2909,  1994.}
\lref\gwg{G.W. Gibbons,
in {\it Fields and Geometry}, proceedings of
22nd Karpacz Winter School of Theoretical Physics: Fields and
Geometry, Karpacz, Poland, Feb 17 - Mar 1, 1986, ed. A. Jadczyk (World
Scientific, 1986);
D. Garfinkle and A. Strominger, \jnl \pl, B256, 146, 1991;
H. F. Dowker, J. P. Gauntlett, S. B. Giddings and G. T. Horowitz,
\jnl \pr, D50, 2662, 1994.}
\lref\ernst{F. J. Ernst, \jnl J. Math. Phys., 17, 515, 1976.}
\lref\schwinger{J. Schwinger, \jnl \pr, 82, 664, 1951.}
\lref\gm{G.W. Gibbons and K. Maeda,
\jnl \np,  B298, 741, 1988.}
\lref\melvin{M. A. Melvin, \jnl \pl, 8, 65, 1964.}
\lref\afma{I.K. Affleck and N.S. Manton, \jnl \np, B194, 38, 1982;
I.K. Affleck, O. Alvarez, and N.S. Manton, \jnl \np, B197, 509, 1982.}
\lref\wald{ R. Wald, {\it General Relativity}, Appendix E,
University of Chicago
Press, 1984.}
\lref\teit{M. Banados, C. Teitelboim, and J. Zanelli, \jnl \prl, 72, 957,
1994.}
\lref\rt{T. Regge and C. Teitelboim, \jnl Ann. Phys., 88, 286, 1974.}
\lref\ad{L. Abbott and S. Deser, \jnl \np, B195, 76, 1982.}
\lref\adm{R. Arnowitt, S. Deser, and C. Misner, in {\it Gravitation:
An Introduction to Current Research}, ed. L. Witten (New York, Wiley) 1962.}
\lref\by{J. D. Brown and J. W. York, {\sl Phys. Rev.} {\bf D47} (1993) 1407.}
\lref\bbwy{ H. Braden, J. Brown, B. Whiting, and J. York, \jnl \pr, D42, 3376,
1990.}
\lref\bcm{J.  D.  Brown, J.  Creighton and R.  B.  Mann, ``Temperature,
Energy and Heat Capacity
of Anti-De Sitter Black Holes", gr-qc 9405007.}
\lref\bmy{See, e.g.,
J. D. Brown, E. A. Martinez, and J. W. York, \jnl \prl, 66, 2281,
1991.}
\lref\am{A. Ashtekar and A. Magnon, \jnl \cqg, 1, L39, 1984.}
\lref\htm{M. Henneaux and C. Teitelboim, \jnl \cmp, 98, 391, 1985.}
\lref\te{ C. Teitelboim, ``Action and Entropy of Extreme and Nonextreme
Black Holes", hep-th/9410103.}
\lref\gk{G. Gibbons and R. Kallosh, ``Topology, Entropy, and Witten
Index of Dilaton Black Holes", hep-th/9407118.}
\lref\vil{A. Vilenkin, \jnl Phys. Rep., 121, 265, 1985.}

\newsec{Introduction}

Traditionally, the gravitational hamiltonian has been studied in the context
of either spatially closed universes or asymptotically flat spacetimes (see
e.g. \rt). In
the latter case, the effect of black hole horizons has been investigated \bmy.
However in recent years, there has been interest in
more general boundary conditions. One example involves the possibility of
a negative cosmological
constant, resulting in spacetimes which asymptotically approach anti-de Sitter
space. Perhaps of greater interest is the
study of the pair creation of black holes in a background magnetic field \gwg.
This involves spacetimes such as the Ernst solution \ernst\
which asymptotically approach the Melvin metric \melvin,
and have a noncompact acceleration horizon  as well as the familiar black
hole horizons. We will give a general derivation of the
gravitational hamiltonian
which can be applied to all spacetimes regardless of their asymptotic
behavior or type of horizons.

In most field theories, the hamiltonian can be derived from the covariant
action
in a straightforward way. In general relativity the situation is complicated
by the fact that the Einstein-Hilbert
action includes a surface term. In most derivations
of the gravitational hamiltonian, the surface term is ignored. This results
in a hamiltonian which is just a multiple of a constraint. One must then add
to this constraint appropriate
surface terms so that its variation is well defined \rt. We will show that
the boundary terms in $H$
come directly from
the boundary terms in the action, and do not need to be added ``by hand".

Since the value of the hamiltonian on a solution is the total energy,
we obtain a definition of the total energy for spacetimes with general
asymptotic behavior. We will show that this definition agrees with
previous definitions in special cases where they are defined. In particular,
for asymptotically flat spacetimes, the energy agrees with the usual
ADM definition \adm, and for asymptotically anti-de Sitter spacetimes
it agrees with the definition proposed by Abbott and Deser \ad.

The relation between the action and the hamiltonian is of special
interest in the euclidean context where it is related to thermodynamic
properties of the spacetime.
For an ordinary field theory, the euclidean action for a static
configuration whose imaginary time is identified with period $\beta$ is
simply $\I = \b H$. It is well known that in general relativity, if there
is a (nonextreme)
black hole horizon present, this relation is modified to  include a
factor of one quarter of the area of the horizon on the right hand
side.
It is clear that an acceleration horizon
must enter this formula differently,  since its area is infinite.
We will derive the general relation  between the euclidean action and
the hamiltonian which applies to acceleration horizons as well as black
hole horizons.

The fact that the naive relation $\I = \b H$  can be modified by black holes
leads to a simple
argument that  the entropy of nonextremal black holes is $S = A/4$, where
$A$ is the horizon area \gh.
It has recently been shown \hhr\ that a similar argument applied directly to
extreme \RN\ black holes yields $S=0$, even though the horizon area is
nonzero (see also \refs{\te,\gk}). We will argue that acceleration horizons
are similar to extreme horizons
in that they also do not contribute to the total entropy, although
for a different reason.

We begin in section 2 by deriving the canonical hamiltonian from the covariant
Einstein-Hilbert action, keeping track of all surface terms. This discussion
applies to spacetimes that can be foliated by complete, nonintersecting
spacelike surfaces. Thus, there are no inner boundaries, and horizons play
no special role at this point. In section 3 we show that the surface term
that arises in the hamiltonian is a reasonable definition of the total energy
for a general spacetime: It agrees with previous definitions when they are
defined. In section 4 we consider the effect of horizons, and derive
the general relation between the hamiltonian and the euclidean action.
We then discuss the entropy, and point out the differences between
horizons of finite and infinite area.


\newsec{Derivation of the Hamiltonian: No inner boundaries}

\subsec{The action}

We start  with the covariant Lorentzian action for a metric $g$ and generic
matter fields $\phi$:
\eqn\action{  I(g,\phi) = \int_M \[ {R\over 16\pi}  +L_m(g, \phi)\]
+{1\over 8\pi}\oint_{\p M} K}
where $R$ is the scalar curvature of $g$, $L_m$ is the matter
lagrangian, and $K$ is the trace of the extrinsic
curvature of the boundary. The surface term is required so that the
action  yields the correct equations of motion subject only to the
condition that the induced three metric and matter fields
on the boundary are held fixed.
(We assume that $L_m$ includes at most first order derivatives.)
The action  \action\
is well defined for spatially compact geometries, but diverges
for noncompact ones. To define the the action for noncompact geometries,
one must choose a reference background $g_0, \phi_0$. We require that this
background be a static solution to the field equations.
The physical action is then the difference
\eqn\physact{  I_P (g, \phi) \equiv I(g,\phi) - I(g_0,\phi_0) \ , }
so the physical action of the reference background is defined to be zero.
$I_P$ is finite for a class of fields $g,\phi$ which asymptotically
approach
$g_0,\phi_0$ in the following sense. We fix a boundary near infinity
$\S^\infty$,
and require that $g,\phi$ induce the
same fields on this boundary as  $g_0, \phi_0$\foot{This condition
can be weakened
so that the induced fields agree to sufficient order so that their
difference does not contribute to the action in the limit that $\S^\infty$
recedes to
infinity.}.

For asymptotically flat spacetimes, the appropriate background is flat space
with zero matter fields,
and \physact\ reduces to the familiar
form of the gravitational action
\eqn\oldact{  I_P(g, \phi) = \int_M \[ {R\over 16\pi}  + L_m\]
  +{1\over 8\pi}\oint_{\p M}
   (K - K_0) }
where $K_0$ is the trace of the extrinsic curvature of the boundary
embedded in flat spacetime.  However, when matter (or a cosmological constant)
is included, one may wish to consider spacetimes which are not asymptotically
flat. In this case one cannot use flat space as the background, and one must
use the more general form of the action \physact.

\subsec{The hamiltonian}

Since the physical action is given by \physact,
the physical hamiltonian is the difference between the hamiltonian computed
from \action\ and the one computed for the background.
To cast the action \action\ into hamiltonian form we follow the discussion
in \wald\ except that all surface terms are retained. To begin, we introduce
 a family of spacelike surfaces $\S_t$ labeled by $t$,  and a timelike
 vector field $t^\mu$ satisfying $t^\mu \d_\mu t =1$. In terms of the
unit normal
$n^\mu$ to the surfaces, we can decompose $t^\mu$ into the usual lapse function
and shift vector
$t^\mu = N n^\mu + N^\mu$. In this section we assume that there are no inner
boundaries, so the
surfaces $\S_t$ do not intersect and  are complete. This does not
rule out the existence of horizons, but it implies that if horizons
form, one continues to evolve the spacetime inside the horizon as well as
outside.
It is convenient to choose the surfaces $\S_t$ so that they meet
the boundary near infinity $\S^\infty$
orthogonally. (This is not essential, but it simplifies the analysis.
Notice that we do not require that $t^\mu$ be tangent to $\S^\infty$.)
Thus the boundary $\p M$ consists
of an initial and final surface with unit normal $n^\mu$, and a surface
near infinity $\S^\infty$ on which $n^\mu$ is tangent.

The four dimensional scalar curvature can be related to
the three dimensional one $\R$ and the extrinsic curvature $K_{\mu\nu}$
of the surfaces $\S_t$ by writing
\eqn\reduce{ R = 2(G_{\mu\nu} - R_{\mu\nu} ) n^\mu n^\nu \ .}
 From the usual initial value constraints, the first term can be expressed
\eqn\constr{ 2G_{\mu\nu} n^\mu n^\nu = \R - K_{\mu\nu} K^{\mu\nu} +K^2 \ .}
The second term can be evaluated by commuting covariant derivatives on $n^\mu$
with the result
\eqn\rterm{ R_{\mu\nu} n^\mu n^\nu = K^2 - K_{\mu\nu} K^{\mu\nu}
     -\d_\mu(n^\mu\d_\nu n^\nu) + \d_\nu(n^\mu\d_\mu n^\nu)  }
When substituted into the action \action, the two total derivative terms
in \rterm\ give rise to boundary contributions. The first is proportional
to $n^\mu$ and hence contributes only on the initial and final boundary.
It completely cancels the  $\oint K$ term on these surfaces. The
second term is orthogonal to $n^\mu$ and only contributes to the surface
integral
near  infinity. If $r^\mu$ is the unit normal to $\S^\infty$, then the
integral over this surface becomes
\eqn\surint{ {1\over 8\pi} \int_{\S^\infty} \d_\mu r^\mu + r_\nu n^\mu \d_\mu
  n^\nu = {1\over 8\pi} \int_{\S^\infty} (g^{\mu\nu} - n^\mu n^\nu) \d_\mu
r_\nu}
This surface integral
has a simple geometric interpretation. The surface $\S^\infty$ is
foliated by a family of two surfaces $S^\infty_t$ coming from its intersection
with  $\S_t$. The integrand in \surint\
is simply the trace
of the two dimensional extrinsic curvature ${}^2 K $ of $S^\infty_t$ in $\S_t$.
Thus the action \action\ takes the form
\eqn\actpart{ I = \int N dt \[{1\over 16 \pi} \int_{\S^\infty} \sqrt { {}^3 g}
   ( \R + K_{\mu\nu} K^{\mu\nu}
  - K^2 +16\pi  L_m) + {1\over 8\pi}\int_{S^\infty_t} {}^2 K \]}
where ${}^3 g$ is the induced metric on $\S_t$.

We now introduce the canonical momenta
$p^{\mu\nu}, p$ conjugate to ${}^3 g_{\mu\nu}, \phi$
and rewrite the action in hamiltonian form. We first consider the
case when the matter does not contain gauge fields. Since the
extrinsic curvature $K_{\mu\nu}$ is related to the time derivative of the
three metric ${}^3 \dot g_{\mu\nu}$ by
\eqn\extcur{ K_{\mu\nu} = {1\over 2N} [  {}^3 \dot g_{\mu\nu} - 2D_{(\mu}
N_{\nu)}] }
where $D_\mu$ is the covariant derivative associated with ${}^3 g_{\mu\nu}$,
when we write the action in a form that does not contain derivatives of
the shift vector, we
obtain another surface term
$-2\int_{S^\infty_t} N^\mu p_{\mu\nu} r^\nu $.
So the action takes the form
\eqn\hact{  I= \int dt \[\int_{\Sigma_t}
  (p^{\mu\nu}\   {}^3 \dot g_{\mu\nu} +p \dot \phi - N \H - N^\mu \H_\mu)
  + {1\over 8\pi}
\int_{S^\infty_t}
 (N\ {}^2 K - N^\mu p_{\mu\nu}r^\nu) \]}
where $\H$ is the Hamiltonian constraint, and $\H_\mu$ is the momentum
constraint. Both of these constraints contain contributions from the matter
as well as the gravitational field.
The hamiltonian  is thus
\eqn\hamil{ H = \int_{\Sigma_t}
   (N\H +N^\mu \H_\mu) - {1\over 8\pi} \int_{S^\infty_t} (N\ {}^2 K -
   N^\mu p_{\mu\nu}r^\nu) \ .}

This expression for the hamiltonian diverges in general,
but recall that physically
we are not interested in the action \action\ but in \physact.
We must therefore derive
the hamiltonian for the reference background. Since this background
is a stationary solution to the field equations, when we repeat the above
analysis using the stationary slices we find that the momenta $p_0^{\mu\nu},
p_0$ vanish and the constraints vanish. If we label the static slices so that
$N_0 = N$ on $\S^\infty$, the reference hamiltonian is simply
\eqn\refham{ H_0 = - {1\over 8\pi} \int_{S^\infty_t} N\ {}^2 K_0 }
  The physical Hamiltonian is the difference
\eqn\physham{  H_P \equiv H - H_0 = \int_{\Sigma_t}
   (N\H + N^\mu \H_\mu) - {1\over 8\pi} \int_{S^\infty_t}\[ N({}^2 K - {}^2
K_0)
    -N^\mu p_{\mu\nu} r^\nu\] }

Given a solution, one can define its total
energy
associated with the time translation $t^\mu = N n^\mu + N^\mu$ to
be simply the value of the physical hamiltonian\foot{Choosing $N=1$ and
$N^\mu =0$, our expression is similar to the one proposed  in \by\ for a
quasilocal energy. However,
the choice of reference background seems highly
ambiguous for a general finite two sphere, while
it is fixed in our approach from the
beginning by the asymptotic behavior of the fields.}
\eqn\energy { E =-{1\over 8\pi} \int_{S^\infty_t} \[N({}^2 K - {}^2 K_0)
    -N^\mu p_{\mu\nu} r^\nu \]}
Notice that the energy of the reference background is automatically zero.
In the next section we will show that  \energy\ agrees with previous
definitions of the energy in special cases where they have been defined.

There is a well known generalization of the above discussion to the case where
the
matter lagrangian contains gauge fields. For example, suppose we start with
the Maxwell lagrangian $L_M = -{1\over 16\pi} F^2 $ where $F=dA$ is the
Maxwell field. Then the canonical variables are the spatial components of
$A_\mu$ and their conjugate momenta $E^\mu$, while the time component $A_t$
acts like a Lagrange multiplier. Using the fact that the inverse spacetime
metric can be
written $g^{\mu\nu} = {}^3 g^{\mu\nu} - n^\mu n^\nu$  with  $n^\mu = (t^\mu -
N^\mu)/N$  one can rewrite the Maxwell action in Hamiltonian form.
The usual energy density ${1 \over 8\pi} (E^2 + B^2)$ is multiplied by the
lapse $N$ and contributes to the Hamiltonian constraint  $\H$.
The usual momentum
density ${1\over 4\pi} \epsilon_{\mu\nu\rho\sigma} n^\nu E^\rho B^\sigma$ is
multiplied by the shift $N^\mu$ and contributes to the momentum constraint
$\H_\mu$. The net result is that the Hamiltonian for the combined
Einstein-Maxwell theory again takes the form \hamil\ except for
an additional
term ${1\over 4\pi} E^\mu D_\mu A_t$ in the volume integral.
This can be integrated by parts to yield
$-A_t/4\pi$ times the
Gauss constraint, $D_\mu E^\mu =0$,
and another surface term ${1\over 4 \pi} \oint_{S_t^\infty} A_t E^\mu r_\mu$.
This term
vanishes for asymptotically flat spacetimes without horizons and for any
purely magnetic field configuration, but it may be nonzero in general.
We shall ignore it in this paper but it is important in electrically charged
black holes \hr.


\newsec{Agreement  with previous expressions for the total energy}

\subsec{Asymptotically flat spacetimes}

In this section we show that the expression for the
total energy obtained  directly from the action in the previous
section \energy\ agrees with  earlier expressions whenever they are defined.
We first consider asymptotically flat spacetimes. Here, the ADM energy
is given by
\eqn\admegy{E_{ADM} = {1\over 16\pi} \oint_{S}( D^i h_{ij} -
       D_j h) r^j}
where the indices $i,j$ run over the three spatial dimensions,
$h_{ij} = {}^3 g_{ij} - {}^3 g_{0ij}$ (${}^3 g_{0ij}$
being the background three-metric \foot{For asymptotically flat spacetimes,
the background three-metric is usually chosen to be flat,
but for later applications
it is convenient to keep the notation general.}),
$D_i$ is the background
covariant derivative,
and $r^i$ is the
unit normal to the large sphere $S$. The energy obtained from the action
\energy\ depends on a choice of lapse and shift. Taking $N=1$ and $N^\mu=0$
(which is appropriate for a unit time translation) yields
\eqn\energ { E =-{1\over 8\pi} \int_{S} ({}^2 K - {}^2 K_0) }
Both \energ\
and \admegy\ are coordinate invariant but depend on a choice of reference
background. We want to show that they are equal whenever the induced
metrics on $S$ agree.\foot{This was also noted in \bbwy.}
 To this end, it is convenient to choose a particular
set of coordinates.
Given a large sphere $S$ in the original spacetime, one can choose coordinates
in a neighborhood of $S$ so that the metric ${}^3 g$ is
\eqn\metr{ ds^2 = dr^2 + q_{ab} dx^a dx^b}
where $a,b$ run over the two angular variables, $r=0$ on $S$,
and the two dimensional metric
$q_{ab}$ is a function of $r$ and $x^a$. Similarly, for the background metric
we can choose coordinates in a neighborhood of $S$ so that the metric
${}^3 g_0$ is
\eqn\bmetr{ ds^2 = d\r^2 + q_{0ab} dy^a dy^b}
We now choose a diffeomorphism from the original spacetime to the background
so that $r=\r, x^a = y^a$. This identification insures that the unit
normal to $S$ in the two metrics agree. Since we are assuming the intrinsic
metric also agrees, $h_{ab} = q_{ab} - q_{0ab} =0$ on $S$.

In these coordinates, we have
\eqn\twok{ \oint_S {}^2 K = {1\over 2} \oint_S q^{ab} (q_{ab,r}) }
So
\eqn\twokd{E= -{1\over 8 \pi}  \oint_S ({}^2 K - {}^2 K_0) =
  - {1\over 16\pi} \oint_S q^{ab} (h_{ab,r}) }
In the ADM  expression \admegy, the first term  can be written
$r^j D^i h_{ij}= D^i( r^j h_{ij}) - h_{ij} D^i r^j$.
The first term on the right is
zero since $h_{ij}$ is always orthogonal to $r^j$,
and the second term is zero since $h_{ij}$ vanishes on $S$. So
\eqn\eadm{ E_{ADM} = -{1\over 16\pi}\oint_S h_{,r} = -{1\over 16\pi}\oint_S
q^{ab} (h_{ab,r})}
where we have again used the fact that $h_{ij}$ vanishes on $S$. Comparing
\twokd\ and \eadm\ we see that the two expressions for the total energy are
equal in this case.

For asymptotically flat spacetimes, one can also define a total momentum.
By taking constant lapse and shift in \energy\ and considering how the
energy changes under boosts of $t^\mu$, one can read off the momentum
\eqn\momen{ P_i N^i = {1\over 8\pi} \oint_S p_{ij} N^i r^j}
which again agrees with the standard ADM result.

\subsec{Asymptotically anti-de Sitter spacetimes}

Abbott and Deser \ad\
have given a definition of the total energy for spacetimes
which asymptotically approach a static solution to Einstein's equation with
negative cosmological constant (see also \refs{\am,\htm}).
If $g_0$ is the static background with timelike Killing
vector $\xi^\mu$, and $h = g-g_0$,
then their definition of the energy is
\eqn\adegy{ E_{AD} = {1\over 8\pi}\oint_S dS_\alpha n_\mu
   [\xi_\nu D_\beta K^{\mu\alpha\nu\beta}
  -K^{\mu\beta\nu \alpha}D_\beta \xi_\nu]   }
where
\eqn\defk{ K^{\mu\alpha\nu\beta} \equiv   g_0^{\mu[\beta} H^{\nu]\alpha}
- g_0^{\alpha[\beta}H^{\nu]\mu} }
 and
\eqn\defH{ H^{\mu\nu} \equiv h^{\mu\nu} - {1\over 2} g_0^{\mu\nu}
h^\alpha_\alpha}
We will again show that $E_{AD}$ agrees with the energy derived from the action
\energy\  when the induced metrics on the surface $S$ agree.
Choosing synchronous gauge for both the physical metric and the background
insures that $h_{0\mu} =0$.
In the spatial gauge described above, $h_{ij} = 0$ on $S$ which implies
$ K^{\mu\alpha\nu\beta} =0$ on $S$, so the second term
in \adegy\ vanishes. If we choose the surface near infinity so that
$\xi^\mu = N n^\mu$, then the first term reduces to
\eqn\ads{E_{AD} = {1\over 16\pi} \oint_{S} N ( D^i h_{ij} -
       D_j h) r^j}
In other words, it is identical to the usual ADM expression except that the
background metric is not flat and the lapse is not one. Since the above
comparison between the ADM expression and \energy\ did not use any special
properties of the flat background and did not involve integration by parts on
the two sphere, it can be repeated in the present context to show
that \ads\ agrees with \energy\ for general
lapse $N$ (and $N^\mu = 0$).
It also agrees with the limit of the quasilocal mass considered in \bcm.

\subsec{Asymptotically conical spacetimes}

As a final comparison of our formula for the energy we consider the
energy per unit length of a cosmic string.\foot{We thank J. Traschen and
D. Kastor for suggesting this example.} Outside the string, the
spacetime takes the form of Minkowski space minus a wedge
\eqn\cosstr{ ds^2 = - dt^2 +dz^2+ dr^2 + a^2 r^2 d\vp^2}
where $\vp$ has period $2\pi$  and the deficit angle is $2\pi ( 1-a)$.
The reference background is flat spacetime without a wedge removed
\eqn\cosstr{ ds^2 = - dt^2 +dz^2+ d\r^2 +  \r^2 d\vp^2}
Since we are interested in the energy per unit
length, we consider a large cylinder  at $r=r_o$ in the cosmic string
spacetime. To match the intrinsic geometry, the corresponding cylinder
in the background has $\r = \r_o$ where $\r_o = ar_o$. The extrinsic curvatures
are ${}^2 K = 1/r_o$ and ${}^2 K_0 = 1/\r_o$. Taking $N=1$ and $N^\mu=0$
in \energy\ yields
\eqn\twokdd{E= -{1\over 8 \pi}  \int ({}^2 K - {}^2 K_0) = -{1\over 8 \pi}
   L  \int \[{1\over r_o} - {1\over \r_o}\] \r_o d\vp = {L\over 4} (1-a) }
  where $L$ is the length of the cylinder.
So $E/L = (1-a)/4$, which agrees with the standard result that the energy
per unit length is equal to the deficit angle divided by $8\pi$ \vil.

\newsec{Horizons and the Euclidean Action }

In section 2 we considered the case where the only boundary of
the surfaces $\Sigma _t$ was at infinity.  However one often
has to deal with cases where the surfaces have an inner boundary as well.
We shall consider two situations:

\item 1. The surfaces $\Sigma _t$ all intersect on a two surface $S_h$.

\item 2. The surfaces $\Sigma _t$ have an internal infinity.
In this case one has to introduce another asymptotic boundary
surface $\Sigma ^{-\infty}$.

The first case arises in spacetimes  containing
a bifurcate Killing horizon, when the surfaces $\Sigma _t$ are adapted to
the time translation symmetry. The
second case arises both for an
extreme horizon, where the intersection between the past and future horizons
has receded to an internal infinity, or for spacetimes having more than
one asymptotic region (such as the maximally extended Schwarzschild solution).
Since we are using a form of the action that requires the metric and
matter fields
to be fixed on the boundary, we shall take them to be fixed on $S_h$ and
$\Sigma ^{-\infty}$.
\foot{We do not require that the fields on $S_h$ agree with those in
the background solution.  Indeed in many cases the background solution
will not possess a two surface of intersection $S_h$. Similarly, for
an internal infinity with finite total action, e.g. resulting from the fact
that the time difference
between the initial and final surface decreases to zero as one moves along an
infinite throat (as in extreme \RN), the background
need not contain an analogous surface $\Sigma ^{-\infty}$. However,
in cases where the internal infinity has infinite action, the background
solution must also contain a surface $\Sigma ^{-\infty} $ on which
the fields agree.}

We shall consider first case (1) where the surfaces of constant time
all intersect on a two surface
$S_h$.  The lapse will be zero on $S_h$
which will be an inner boundary to the surfaces
$\Sigma _t$.  We  can also
choose the shift vector to vanish on this boundary. One
can now repeat the derivation of the hamiltonian given in section 2.
The only difference
is that the surface term ${1\over 8\pi}
\oint N\ {}^2 K$ will now appear on the inner
boundary as
well as at infinity. However, this term
vanishes since the lapse $N$ goes to zero
at $S_h$.
If the reference background also has a horizon, there
will be an extra surface term  ${1\over 8\pi}\oint N_0\ {}^2K_0$
coming from the inner boundary
there. But this will also vanish since $N_0$ vanishes at the
horizon. Thus the
hamiltonian generating evolution outside a horizon $S_h$
is again given by \hamil\ with only a surface term at infinity.\foot{If one
does
not keep the metric on the boundary fixed, the hamiltonian picks up a surface
term proportional to the derivative of the lapse \bmy.}

If the surfaces $\Sigma _t$ do not intersect
but have an internal infinity, there will be
a surface term ${1\over 8\pi} \oint N\ {} ^2K$
on $\Sigma ^{-\infty} $.  For spacetimes like extreme \RN\
this will be zero because ${} ^2K$
will go to zero as one goes down the throat, as will the lapse
$N$ corresponding to the time translation Killing vector.
However in the case of the maximally extended Schwarzschild solution, the
surface term (including the background contribution) is ${1\over 8\pi}
\oint N({} ^2K - {} ^2K_0)$ which can
contribute to the value
of the hamiltonian.

We now consider the euclidean action
\eqn\eaction{\I = -{1\over 16 \pi} \int_M (R + 16\pi L_m) -{1\over
8\pi}\oint_{\p M} K}
In a static or stationary solution the time derivatives $({}^3\dot g_{\mu\nu},
\dot \phi )$
are zero.
Thus the action for a region between surfaces
$\Sigma _t$ an imaginary time distance $\b$ apart is
\eqn\hami{\I =\b H}
If the stationary time surfaces $\Sigma _t$
do not intersect, then the
imaginary time coordinate
can be periodically identified with any period $\b$.
This is the case for the extreme  Reissner-Nordstr\"om
black hole since the horizon is infinitely far away.
For such periodically identified solutions,
the total
action will be given by \hami.  However when the stationary time surfaces
intersect at a horizon $S_h$,  the periodicity $\b$ is fixed by regularity
of the euclidean solution at $S_h$.
The action
of the region swept out by the surfaces $\Sigma_t$ between their inner
and outer boundaries is again $\I =\b H$.  However this is not the action of
the full four dimensional solution \hawk, but only of the solution with the
two surface $S_h$ removed.  The
contribution to the action from a little tubular neighborhood
surrounding the two surface $S_h$ is just $-A/4$ (see also \teit) where $A$ is
the area of $S_h$. We thus  obtain
\eqn\hrrel{ \I = \b H - {1\over 4} A}

As they stand, \hami\ and \hrrel\ are meaningless since we have not yet
taken into account the reference background. Consider first the case where the
background
does not contain a two surface $S_h$ on which the stationary time surfaces
intersect.  The background must be identified with the same period in imaginary
time
at infinity as the solution under consideration
in order for the induced metrics
on $\S^\infty$ to agree. One thus obtains $\I_0 = \b H_0$ for the
background which leads to the familiar result
\eqn\famil{ \I_P = \b H_P - {1\over 4} A_{bh}}
for the case of nonextreme black holes but
\eqn\ifamil{\I_P=\b H_P}
in the extreme case.

As is now well known \gh, the path integral
over all euclidean metrics and matter fields that are
periodic with period $\b$ at infinity gives
the partition function at temperature $T=\b^{-1} $
\eqn\part{Z=\sum _{\rm states} e^{-\b E_n} =
\int D[g] D[\phi] e^{-\I_P}}
In the semiclassical approximation, the dominant
contribution to the path integral will come from the neighborhood of
saddle points of the action, that is, of classical solutions.  The
zeroth order contribution to $\log Z$ will be $-\I _P$.
All thermodynamic properties can be deduced from the partition function.  For
instance, the expectation value of the energy is
\eqn\en{\langle E \rangle =-{\p \over \p \b} \log Z}
By \famil\ or \ifamil\ the zeroth order contribution to $\langle E \rangle $
will be $H_P$, as one might expect.  The entropy can be defined by
\eqn\ent{S=-\sum p_n \log p_n=- \( \b {\p \over \p \b} -1\) \log Z}
where $p_n=Z^{-1}e^{-\b E_n} $ is the probability of being in the nth state.
If one applies this to the expressions for the action \famil\ and \ifamil,
one sees that
the zeroth order contribution to the entropy
of an extreme black hole is zero \hhr.  On the other
hand, the entropy of a nonextreme
black hole is $A_{bh}/4$.

So far  we have assumed implicitly that the horizon two surface $S_h$
is compact so that its area is finite.  We now consider
the case when the area of $S_h$ is infinite, such as for
acceleration horizons.  The main difference between this case and the previous
one comes from the fact that the horizon now extends out to infinity. One
could try to keep
the surface $\S^\infty$ away from the horizon, but then the
space between $\S^\infty$  and the horizon would still be
noncompact, so  the action would be ill-defined. If the spacetime has
continuous spacelike symmetries, one could compute all quantities per unit
area.
Alternatively, if the spacetime has appropriate discrete symmetries,
one could periodically
identify to make the action (and horizon area) finite.
If either of these two options is adopted, then
the previous discussion applies essentially unchanged. However, in
general, neither option is available. One must then
choose $\S^\infty$ to intersect the horizon ``at infinity".
Thus, instead of the intersections of $\S^\infty $ and the surfaces $\S_t$
having topology $S^2 $, they  will now
have topology $D^2$.
Since the metric induced on
$\S^\infty$ from the background spacetime
must agree with that from the original
spacetime,
it follows that the background metric must also have a
horizon that intersects $\S^\infty$.

As a simple example,
consider Rindler space
\eqn\rindler{ds^2 = -\xi^2 d\eta^2 + d\xi^2 +
dy^2 + dz^2}
If one does not periodically identify $y$ and $z$ (or compute
quantities per unit area), one must take
$\S^\infty$ to be given by fixing a large value
of $R^2 = \xi^2 + y^2 + z^2$, which intersects the horizon $\xi=0$.
The surfaces of constant $\eta$ intersect $\S^\infty$ in a disk $D^2$ since
$\xi \ge 0$.

We now consider the euclidean version of solutions with acceleration horizons.
The argument above \hrrel\ can be applied to show that \hrrel\ holds in this
case also.
Since the periodicity in imaginary time
is determined by regularity of the euclidean spacetime on the axis (which
now extends out to infinity) the periodicity in the background $\b_0$
must again agree with that in the original spacetime  $\b$.
Repeating the argument above \hrrel\ one finds that the background
satisfies a similar relation
\eqn\hzrel{ \I_0 = \b H_0 - {1\over 4} A_0}
Thus, the physical euclidean action is related to the physical hamiltonian by
\eqn\genrel{ \I_P = \b H_P - {1\over 4} \Delta A}
where $\Delta A$ is the difference between the area of $S_h$ in the
original spacetime and its area in the reference background. This general
formula
includes the familiar result \famil\ as a special case, since for black hole
horizons, one can choose a background which does not have a horizon.
If several horizons $S_h$
are present, $\Delta A$ is the
increase in area of the acceleration horizon
{\it plus} the area of any
nonextreme black hole horizons.  It does not however include the area of
extreme horizons because they do not meet at a two surface in the spacetime.

Since the area of an acceleration horizon is infinite, one might think
that the difference $\Delta A$ is ill-defined. However, it can be given
a precise meaning by examining how it enters into the above argument. The
main point is that the surface near
infinity $\S^\infty$ intersects the acceleration horizon at a large but finite
circle $C$. $\Delta A$ is defined to be the difference between the (finite)
area of the acceleration horizon inside $C$ in the original spacetime and
the area inside the analogous circle $C_0$
in the reference background. Since the
fields induced on $\S^\infty$ from the original spacetime agree
with those induced from the reference background, one can rephrase this
prescription as follows: One fixes a large circle $C$ in the acceleration
horizon in the original spacetime and then chooses a circle $C_0$
in the reference
background which has the same proper length and the same value of the matter
fields. $\Delta A$ is then the difference in area inside these two circles.
This procedure was used in \hhr\ to analyze the Ernst
instanton.

If one naively substitutes the euclidean action \genrel\ into the expression
for the entropy \ent\ using the zeroth order
contribution $\log Z \approx -\tilde I_P$, one might conclude that
an acceleration horizon should have an entropy $\Delta A/4$.
However,
the periodicity of the
imaginary time coordinate on the boundary is fixed by the
requirement of regularity where the acceleration horizon meets  $\S^{\infty} $.
Thus one cannot take the derivative of the partition function with respect to
 $\b$ and so cannot use \ent\ to calculate the entropy.
 This differs from the black hole case where $\b$ is not fixed by regularity
 at infinity.
Instead, we shall use a different argument. Physically, a key difference
between acceleration and black hole horizons is that the former are observer
dependent. The information behind an acceleration horizon can
be recovered by observers who simply stop accelerating.
Another way to say this is that acceleration horizons are not associated
with a change in the topology of spacetime. For example,
consider a spacetime like
the Ernst solution where there are both acceleration and black hole horizons.
One could imagine replacing the black holes by something like magnetic
monopoles
that have no horizons.  One could make the monopole solution away from the
black hole horizons
arbitrarily close to the solution with black holes.  The monopole solution
would have the same $R^4$ topology as the Melvin reference background.  Thus
one
could choose a different family $\S ^{\prime}_t$ of time surfaces that cover
the
region within a large three sphere without intersections or inner boundaries.
One would therefore expect the monopole solution to have a unitary hamiltonian
evolution
and zero entropy.

However, the area of the acceleration horizon in the monopole solution
will still be different from that of the background.
Since $H_P=0$ \hhr, this difference $\Delta A_{acc}$
is directly related to the euclidean action \genrel\ and
thus will correspond to the tunneling probability to create a
monopole-antimonopole pair (assuming there is only one
species of monopole).
However the instanton representing the pair creation of nonextremal
black holes will have
a lower action because there is an extra contribution to $\Delta A$ from
the black hole horizon area $A_{bh} $.  One can interpret the increased
pair creation probability as corresponding to the possibility of
producing $N=\exp (A_{bh}/4)$ different species of black hole pairs.  Thus
pair creation arguments confirm the connection between entropy and (nonextreme)
black hole horizon area, but suggest that there is no analogous
connection with acceleration horizon area.

\vskip 1cm

\centerline{\bf Acknowledgments}

We wish to thank C. Teitelboim for extensive discussions at the Durham
Symposium on Quantum Concepts in Space and Time, July, 1994.
G. H. was supported in part by NSF grant PHY-9008502.

\vfill\eject

\listrefs

\end